\documentclass[12pt]{article}
\usepackage{epsfig}

\usepackage{amssymb}
\usepackage{amsmath}
\usepackage{amsfonts}

\usepackage{amsthm}

\def\be{\begin{equation}}
\def\ee{\end{equation}}
\def\ba{\begin{array}{c}}
\def\ea{\end{array}}

\begin{document}


\vspace{.35cm}

 \begin{center}{\Large \bf

Complex covariance\footnote{Material presented
during the AAMP 10 conference in Prague\\ (Villa Lanna, June 4 - 7,
2012,
http://gemma.ujf.cas.cz/$\sim$znojil/conf/micromeetingdeset.html)\vspace{2mm}}

  }\end{center}

\vspace{10mm}

 \begin{center}

 {\bf Frieder Kleefeld}\footnote{Present permanent address: Pfisterstr. 31, 90762 F\"urth, Germany}

 \vspace{3mm}
Collaborator of the

Centro de F\'{\i}sica das Interac\c{c}\~{o}es Fundamentais (CFIF), 

Instituto Superior T\'{e}cnico (IST),

Edif\'{\i}cio Ci\^{e}ncia, Piso 3,  Av. Rovisco Pais, 

P-1049-001 LISBOA, Portugal

{e-mail: kleefeld@cfif.ist.utl.pt}



\end{center}


\section*{Abstract}
According to some generalized correspondence principle the classical limit of a non-Hermitian Quantum theory describing quantum degrees of freedom is expected to be well known classical mechanics of classical degrees of freedom in the complex phase space, i.e., some phase space spanned by complex-valued space and momentum coordinates. As special relativity has been developed by Einstein merely for real-valued space-time and four-momentum we will try to understand how special relativity and covariance can be extended to complex-valued space-time and four-momentum. Our considerations will lead us not only to some unconventional derivation of Lorentz transformations for complex-valued velocities, yet also to the non-Hermitian Klein-Gordon and Dirac equations which are to lay the foundations of a non-Hermitian quantum theory.   

\newpage

\section{Introduction \label{I} }
As pointed out at various places (see e.g.\ 
\cite{Kleefeld:2006bp,Kleefeld:2004jb,Kleefeld:2003dx,Kleefeld:2003zj,Kleefeld:2002gw,Kleefeld:2002au,Kleefeld:2001xd,Kleefeld:1999th,Kleefeld:1998dg,Kleefeld:1998yj} and references therein) a simultaneous causal, local, analytic and covariant formulation of physical laws requires a non-Hermitian extension of quantum theory, i.e. quantum mechanics and quantum field theory.
Since causality expressed in quantum theory by the time-ordering operation is inferring some small negative imaginary part in self-energies appearing in causal propagators which may be represented by some (in most cases infinitesimal) negative imaginary part in particle masses, even field operators representing electrically neutral causal particles have to be considered non-Hermitian. This leads to the fact that their creation and annihilation operators are not Hermitian conjugate to each other.  During the attempt to find some spacial representation of non-Hermitian creation and annihilation operators which preserves analyticity and covariance it has turned out that aforementioned non-Hermitian causal field operators are functions of the complex spacial variable $z$ instead of merlely the real spacial variable $x$, while anticausal field operators are functions of the the complex conjugate spacial variable $z^\ast$. Consequently a causal, local, analytic and covariant formulation of the laws of nature separates into some holomorphic causal sector and some antiholomorphic anticausal sector which must not interact on the level of causal and anticausal field-operators in the spacial respresentation.

At least since Gregor Wentzel \cite{Wentzel:1926xy} (1926) there exists a formalism (see also \cite{Yang:2005ub,John:2001zc,Faraggi:1998pd,Leacock:1983zz,Bender:1977dr,Dawydow:1973,Dunham:1932xy}) called nowadays e.g.\ ``Quantum-Hamilton-Jacobi Theory" (QHJT) or ``Modified de Broglie-Bohm Approach" which relates a field or wave function by some correspondence principle (see e.g.\ our Eqs. (\ref{corry})) to the trajectories of some ``quantum particle" in the whole complex phase space. Wentzel's approach has been recently even fortified by A.\ Voros \cite{Voros:2012se} (2012) by providing some ``exact WKB method" allowing to solve the Schr\"odinger equation for arbitrary polynomial potentials simultaneously in the whole complex $z$-plane. Moreover there exists  a rapidly increasing interest \cite{PTconferences} of many theoretical and experimental researchers to study solutions of the Schr\"odinger equation even for non-Hermitian Hamiltonians in the whole complex plane due to a meanwhile confirmed conjecture of D.~Bessis (and J.\ Zinn-Justin) in 1992 on the reality and positivity of spectra for manifestly non-Hermitian Hamiltonians which was related by C.M. Bender and S. Boettcher in 1997 \cite{Bender:1998ke} to the PT-symmetry \cite{Bender:2007nj}
 of these Hamiltonians.

Despite this enormous amount of activities to ``make sense of non-Hermitian Hamiltonians" \cite{Bender:2007nj,Bender:2009wx} and the fact that we had managed \cite{Kleefeld:2002au} to construct even a Lorentz-boost for complex-mass fields required to formulate non-Hermitian spinors and Dirac-equations there has remained to our understanding one crucial point neglected and unclear which is is in the spirit of the QHJT and which will be the focus of the presented results: {\em How does the general concept of ``covariance" extend to complex phase space which has been formulated by Albert Einstein (1905)} \cite{Einstein:1905ve} {\em and collegues (see also} \cite{Pais2000,History2012}) {\em merely for a phase-space in which the spacial and momentum coordinates are considered to be real-valued?} An answer will be given to some extent in the follwoing text.
Certainly one might argue that there exist already approaches like e.g.\ \cite{Nakanishi:1972wx,Nakanishi:1972pt,Lee:1970iw,Mostafazadeh:2006fz,Bender:2011ke} which refer to seemingly covariant equations within a non-Hermitian framework. It turns nonetheless out that in all of the approaches there remain open questions to the reader in how far these equations are consistent with some of the aspects causality, locality or analyticity and in how far the four vectors formed by space-time and momentum-energy coordinates contained in these equations will transform consistently under Lorentz transformations as it seems at the first sight (e.g.\ \cite{Nakanishi:1971jj,Gleeson:1972xj}) completely puzzling how to extend the framework of an inertial frame to the complex plane.

\section{Space-time covariance for complex-valued velocities}

The purpose of this section is the derivation of generalized Lorentz-Larmor-FitzGerald-Voigt (LLFV) \cite{Lorentz:1899xy,Poincare:1905xy} transformations\footnote{Major steps in the LLFV-history: stepwise derivation of the transformations by Voigt (1887), FitzGerald (1889), Lorentz (1895, 1899, 1904), Larmor (1897, 1900); formalistic progress afterwards: Poincar\'{e} (1900, 1905) (discovery of Lorentz-group properties and some invariants), Einstein (1905) (derivation of LLFV transformations from first principles), Minkowski (1907-1908) (geometric interpretation of LLFV transformations).} relating the space-time coordinates $\vec{z},t$ and $\vec{z}^{\,\prime},t^\prime$ of two inertial frames ${\cal S}$ and ${\cal S^\prime}$, respectively, which move with some constant 3-dimensional complex-valued relative velocity $\vec{v}\in \mathbb{C}\times \mathbb{C}\times\mathbb{C}$. While the 3-dimensional space vectors $\vec{z}$ and $\vec{z}^{\,\prime}$ are assumed to be complex-valued, i.e.\  $\vec{z},\vec{z}^{\,\prime}\in \mathbb{C}\times \mathbb{C}\times\mathbb{C}$, our derivation will display how the preferably real-valued time coordinates $t$ and $t^\prime$ will be complexified.

Without loss of generality we will constrain ourselves throughout the derivation to one complex-valued spacial dimension while the generalization to three complex-valued spacial dimensions appears straight forward. Hence we consider in what follows for simplicity  generalized LLFV transformations relating space-time coordinates $z,t$ and $z^\prime,t^\prime$  (with $z,z^\prime\in \mathbb{C}$) of two inertial frames ${\cal S}$ and ${\cal S^\prime}$, respectively, moving with some constant one-dimensional complex-valued relative velocity $v\in \mathbb{C}$.

According to the standard definition an inertial frame in the absence of gravitation is a system in which the first law of Newton holds. For our purposes we will rephrase this definition in a way which may be used even in some complexified space-time: 

{\em An inertial frame in the absence of gravitation is a system whose trajectory in (even complexified) space-time is a straight line.}

Our definition of inertial frames implies directly that generalized LLFV transformations relating inertial frames must be {\em linear}. Making use of this observation we can write down --- as a first step in our derivation --- the following linear ansatz for the  generalized LLFV transformations between the inertial frames ${\cal S}$ and ${\cal S^\prime}$:
\begin{eqnarray} z^\prime & = & \gamma \cdot z + \delta \cdot t + \varepsilon \, , \label{step1a} \\
t^\prime & = & \kappa \cdot z + \mu \cdot t + \nu \, . \label{step1b}
\end{eqnarray}
Here $\gamma$, $\delta$, $\varepsilon$, $\kappa$, $\mu$, $\nu$ are yet unspecified eventually complex-valued constants.

In a second step we will perform --- without loss of generality --- a synchronization of the inertial frames ${\cal S}$ and ${\cal S^\prime}$ by imposing the following condition: 
\begin{eqnarray} (z,t)=(0,0) & \Leftrightarrow & (z^\prime,t^\prime)=(0,0) \, .
\end{eqnarray}
Obviously the synchronisation yields $\varepsilon=\nu=0$. Inserting this result in Eqs. (\ref{step1a}) and (\ref{step1b}) leads to the following equations:
\begin{eqnarray} z^\prime & = & \gamma \cdot z + \delta \cdot t \quad = \quad \gamma \left(  z + \frac{\delta}{\gamma} \cdot t \right)  \, , \label{step2a} \\
t^\prime & = & \kappa \cdot z + \mu \cdot t \, . \label{step2b}
\end{eqnarray}
In the 3rd step we  use of the relative complex-valued velocity between inertial frames  ${\cal S}$ and ${\cal S^\prime}$: the spacial orgin $z^\prime = \gamma \cdot z + \delta \cdot t = 0$ of  ${\cal S^\prime}$ moves in  ${\cal S}$ with constant complex-valued velocity $v\equiv z/t = - \delta/ \gamma$ yielding for Eq. (\ref{step2a}):
\begin{eqnarray} z^\prime & = & \gamma \,  (  z - v \cdot t )  \quad \mbox{with} \quad \gamma = \gamma(v) \, . \label{step3a} 
\end{eqnarray}
In writing $\gamma = \gamma(v)$ we point out that the constant $\gamma$ could be a function of the complex-valued velocity $v$.

The fourth step is the application of the principle of relativity which states that --- in the absence of gravity --- {\em there does not exist any preferred inertial frame of reference implying in particular that the laws of physics take the same mathematical form in all inertial frames}. It provides an essentially unique prescription of how to construct inverse generalized LLFV transformations: interchange the space-time coordinates $z,t$ and $z^\prime, t^\prime$, respectively, and replace the complex-valued velocity $v$ by $-v$. As a consequence we obtain for the corresponding ``inverse" of Eq.\  (\ref{step3a}):
\begin{eqnarray} z & = & \overline{\gamma} \,  \left(  z^\prime + v \cdot t^\prime \, \right)  \quad \mbox{with} \quad \overline{\gamma} \equiv \gamma(-v) \, . \label{step4a} 
\end{eqnarray}
In order to determine the yet unknown eventually complex-valued constants $\kappa$ and $\mu$ in Eq.\ (\ref{step2a}) we solve Eq.\ (\ref{step4a}) for $t^\prime$ and apply to the result the identity Eq.\ (\ref{step3a}), i.e.:
\begin{eqnarray} t^\prime & = & \frac{1}{v} \left( \frac{z}{\overline{\gamma}} - z^\prime \right) \quad = \quad  \frac{1}{v} \left( \frac{z}{\overline{\gamma}} -  \gamma \,  (  z - v \cdot t ) \right)
\end{eqnarray}
or --- after some rearrangement ---
\begin{eqnarray} t^\prime & = & \gamma \,  \left(  t - \frac{1}{v} \left(1 - \frac{1}{\gamma \,\overline{\gamma}} \right) \! z \right)  \, . \label{step4b} 
\end{eqnarray}
Comparison of Eq.\ (\ref{step4b}) with Eq.\ (\ref{step2b}) yields $\kappa = - \frac{\gamma}{v}\left(1 - \frac{1}{\gamma \,\overline{\gamma}} \right)$ and $\mu = \gamma$. By the same procedure leading from Eq.\ (\ref{step3a}) to Eq.\ (\ref{step4a}) the principle of relativity can be used to obtain the corresponding ``inverse" of Eq.\  (\ref{step4b}), i.e.:
\begin{eqnarray} t & = & \overline{\gamma} \,  \left(  t^\prime + \frac{1}{v} \left(1 - \frac{1}{\overline{\gamma}\,\gamma} \right) \! z^\prime \right)  \, . \label{step4c} 
\end{eqnarray}
Hence, the previous considerations result in the following four identities (with $\gamma \equiv \gamma (v)$, $\overline{\gamma} \equiv \gamma (-v)$):
\begin{eqnarray} z^\prime & = & \gamma \,  (  z - v \cdot t )  \, , \label{step4aa} \\[1mm]
 t^\prime & = & \gamma \,  \left(  t - \frac{1}{v} \left(1 - \frac{1}{\gamma \,\overline{\gamma}} \right) \! z \right)  \, , \label{step4bb} \\[3mm]
 z & = & \overline{\gamma} \,  \left(  z^\prime + v \cdot t^\prime \, \right) \, , \label{step4aaa} \\[1mm]
 t & = & \overline{\gamma} \,  \left(  t^\prime + \frac{1}{v}\left(1 - \frac{1}{\overline{\gamma}\,\gamma} \right) \! z^\prime \right)  \, . \label{step4bbb} 
\end{eqnarray}
In dividing Eq.\ (\ref{step4aa}) by Eq.\ (\ref{step4bb}) and  Eq.\ (\ref{step4aaa}) by Eq.\ (\ref{step4bbb}) we obtain a generalized velocity addition law Eq.\ (\ref{step4cc}) and its inverse Eq.\ (\ref{step4dd}), respectively, i.e.:  
\begin{eqnarray} \frac{z^\prime}{t^\prime} & = & \frac{ \displaystyle \frac{z}{t}- v}{\displaystyle 1 - \frac{1}{v} \left(1 - \frac{1}{\gamma \,\overline{\gamma}} \right) \! \frac{z}{t}}  \, , \label{step4cc} \\
 \frac{z}{t} & = & \frac{ \displaystyle \frac{z^\prime}{t^\prime}+ v}{\displaystyle 1 + \frac{1}{v} \left(1 - \frac{1}{\overline{\gamma}\,\gamma} \right) \! \frac{z^\prime}{t^\prime}}  \, . \label{step4dd}
\end{eqnarray}

A final fifth step makes use of the principle of constancy inferred by Albert Einstein stating that {\em light in vacuum is propagating in all inertial frames with the same speed independent of the movement of the light source and the propagation direction.} For our purposes we will generalize and simplify this principle of constancy by simply claiming that {\em the velocity addition law and its inverse possess an eventually complex-valued fixed point $c$ whose modulous $|c|$ coincides with the vacuum speed of light}. Or, in other words: {\em there exists some eventually complex-valued velocity $c$ which is not modified by the application of  the velocity addition law and its inverse while the modulous  $|c|$ coincides with the vacuum speed of light.} Application of this generalized principle of constancy to the addition laws Eq.\ (\ref{step4cc}) and Eq.\ (\ref{step4dd}) yields the following identity Eq.\ (\ref{step5a}):
\begin{eqnarray} c \; = \; \frac{ \displaystyle c\mp v}{\displaystyle 1 \mp \frac{1}{v} \left(1 - \frac{1}{\gamma \,\overline{\gamma}} \right) \! c} \quad & \Longrightarrow &  \quad \gamma \,\overline{\gamma} \; = \; \frac{1}{\displaystyle 1-\frac{v^2}{c^2}} \, . \label{step5a} 
\end{eqnarray}
As $\gamma \,\overline{\gamma}$ depends on $c^2$, i.e., the square of $c$, we can conclude that ---  besides the fixed point $+c$ of the velocity addition law Eq.\ (\ref{step4cc}) and its inverse --- there simultaneously exists a second fixed point $-c$. Eq.\ (\ref{step5a}) can be used to transform Eqs. (\ref{step4aa}),  (\ref{step4bb}), (\ref{step4aaa}), (\ref{step4bbb}) and (\ref{step4cc}), (\ref{step4dd}) to their final form. As LLFV transformations should reduce to the identity in the limit \mbox{$v\rightarrow 0$}, we take the ``positive" complex square root of  Eq.\ (\ref{step5a}), i.e., $(\gamma \,\overline{\gamma})^{-1/2}=+\sqrt{ 1-\frac{v^2}{c^2}}$, and invoke it together with $\gamma = \sqrt{\gamma \, \overline{\gamma}} \cdot \sqrt{\frac{\gamma}{\overline{\gamma}}}$ and $\overline{\gamma} = \sqrt{\gamma \,\overline{\gamma}} \cdot \sqrt{\frac{\overline{\gamma}}{\gamma}}$ to  Eqs. (\ref{step4aa}),  (\ref{step4bb}), (\ref{step4aaa}), (\ref{step4bbb}) and obtain the following generalized  LLFV transformations  (with $\gamma \equiv \gamma (v)$, $\overline{\gamma} \equiv \gamma (-v)$):
\begin{eqnarray} z^\prime & = &  \sqrt{\frac{\gamma}{\overline{\gamma}}}\, \cdot \,  \frac{\displaystyle z - v \cdot t}{\displaystyle \sqrt{ 1-\frac{v^2}{c^2}}}  \, , \label{step6a} \\[1mm]
  t^\prime & = &  \sqrt{\frac{\gamma}{\overline{\gamma}}}\, \cdot \,  \frac{\displaystyle t - \frac{v}{c^2} \cdot z}{\displaystyle \sqrt{ 1-\frac{v^2}{c^2}}}  \, , \label{step6b} \\[3mm]
 z & = &  \sqrt{\frac{\overline{\gamma}}{\gamma}}\, \cdot \,  \frac{\displaystyle z^\prime + v \cdot t^\prime}{\displaystyle \sqrt{ 1-\frac{v^2}{c^2}}}  \, , \label{step6c} \\[1mm]
  t & = &  \sqrt{\frac{\overline{\gamma}}{\gamma}}\, \cdot \,  \frac{\displaystyle t^\prime + \frac{v}{c^2} \cdot z^\prime}{\displaystyle \sqrt{ 1-\frac{v^2}{c^2}}}  \, , \label{step6d}
\end{eqnarray}
and to Eqs.\ (\ref{step4cc}), (\ref{step4dd}) to arrive at the generalized velocity addition law and its inverse:
\begin{eqnarray} \frac{z^\prime}{t^\prime} & = & \frac{ \displaystyle \frac{z}{t}- v}{\displaystyle 1 - \frac{v}{c^2}  \cdot \frac{z}{t}}  \, , \label{step6e} \\
 \frac{z}{t} & = & \frac{ \displaystyle \frac{z^\prime}{t^\prime}+ v}{\displaystyle 1 + \frac{v}{c^2} \cdot \frac{z^\prime}{t^\prime}}  \, . \label{step6f}
\end{eqnarray}
Two comments are here in order: even though the previous Eqs.\ (\ref{step6a}), (\ref{step6b}), (\ref{step6c}), (\ref{step6d}), (\ref{step6e}) and (\ref{step6f}) look very similar to the well known text book equations appearing in the context of the standard formalism of special relativity they are completely non-trivial as they hold not only in a real-valued space-time and for real-valued velocities, yet also {\em in a complex-valued space-time and for complex-valued velocities}. Moreover is the extension of the aforementioned equations to three spacial dimensions achieved by replacing the complex-valued quantities $z$, $z^\prime$ and $v$ by complex-valued 3-dimensional vectors  $\vec{z}$, $\vec{z}^{\,\prime}$ and $\vec{v}$, respectively.
\section{On the choice of the invariant velocities $\pm c$ and the complexification of time}
As we allow complex-valued velocities we face more freedom than Albert Einstein in defining the invariant eventually complex-valued invariant velocities $\pm c$. We will discuss here two specific options to define $c$ of which the first is our preferred choice due to the arguments given below:\\
{\bf Option 1:} {\em Choose $c = \pm |c|$ real-valued with $|c|=299792458$~{\rm m/s}} \cite{mohr2012} {\em being the vacuum speed of light and set $\gamma = \overline{\gamma}$.}\\
Performing this choice  Eqs.\ (\ref{step6a}), (\ref{step6b}), (\ref{step6c}), (\ref{step6d}), (\ref{step6e}) and (\ref{step6f}) read:
\begin{eqnarray} z^\prime & = &   \frac{\displaystyle z - v \cdot t}{\displaystyle \sqrt{ 1-\frac{v^2}{|c|^2}}}  \, , \qquad   t^\prime \; = \;  \frac{\displaystyle t - \frac{v}{|c|^2} \cdot z}{\displaystyle \sqrt{ 1-\frac{v^2}{|c|^2}}}  \, ,  \qquad \frac{z^\prime}{t^\prime} \; = \; \frac{ \displaystyle \frac{z}{t}- v}{\displaystyle 1 - \frac{v}{|c|^2}  \cdot \frac{z}{t}}  \, ,  \\
 z & = &  \frac{\displaystyle z^\prime + v \cdot t^\prime}{\displaystyle \sqrt{ 1-\frac{v^2}{|c|^2}}}  \, , \qquad t \; = \;   \frac{\displaystyle t^\prime + \frac{v}{|c|^2} \cdot z^\prime}{\displaystyle \sqrt{ 1-\frac{v^2}{|c|^2}}}  \, , \qquad \frac{z}{t} \; = \; \frac{ \displaystyle \frac{z^\prime}{t^\prime}+ v}{\displaystyle 1 + \frac{v}{|c|^2} \cdot \frac{z^\prime}{t^\prime}}  \, . 
\end{eqnarray}
With the exception of the square-roots all these equations are manifestly analytic.
On the world-line $z=v\cdot t$ of  ${\cal S^\prime}$ in ${\cal S}$ we have with $t\in  \mathbb{R}$:
\begin{eqnarray}  t^\prime & = &  \frac{\displaystyle t - \frac{v}{|c|^2} \cdot v\cdot t}{\displaystyle \sqrt{ 1-\frac{v^2}{|c|^2}}} \; = \;  t\cdot  \sqrt{ 1-\frac{v^2}{|c|^2}} \; \in \mathbb{C} \quad \mbox{for} \quad v\in \mathbb{C} \; .
\end{eqnarray}
Hence, the attractive feature of analyticity would be obtained at the price of multiplying time in the boosted frame by some complex-valued constant.\\
{\bf Option 2:} {\em Choose $c$ and $v$ to be (anti)parallel in the complex plane, i.e., $c = \pm |c|\cdot \frac{v}{|v|}=\pm \left|\frac{c}{v}\right|\cdot v$ with $|c|=299792458$~{\rm m/s}}  \cite{mohr2012} {\em being the vacuum speed of light, and set $\gamma = \overline{\gamma}$.}\\
Performing this choice  Eqs.\ (\ref{step6a}), (\ref{step6b}), (\ref{step6c}), (\ref{step6d}), (\ref{step6e}) and (\ref{step6f}) read:
\begin{eqnarray} z^\prime & = &   \frac{\displaystyle z - v \cdot t}{\displaystyle \sqrt{ 1-\left|\frac{v}{c}\right|^2}}  \, , \qquad   t^\prime \; = \;  \frac{\displaystyle t - \frac{v^\ast}{|c|^2} \cdot z}{\displaystyle \sqrt{ 1-\left|\frac{v}{c}\right|^2}}  \, ,  \qquad \frac{z^\prime}{t^\prime} \; = \; \frac{ \displaystyle \frac{z}{t}- v}{\displaystyle 1 - \frac{v^\ast}{|c|^2}  \cdot \frac{z}{t}}  \, ,  \\
 z & = &  \frac{\displaystyle z^\prime + v \cdot t^\prime}{\displaystyle \sqrt{ 1-\left|\frac{v}{c}\right|^2}}  \, , \qquad t \; = \;   \frac{\displaystyle t^\prime + \frac{v^\ast}{|c|^2} \cdot z^\prime}{\displaystyle \sqrt{ 1-\left|\frac{v}{c}\right|^2}}  \, , \qquad \frac{z}{t} \; = \; \frac{ \displaystyle \frac{z^\prime}{t^\prime}+ v}{\displaystyle 1 + \frac{v^\ast}{|c|^2} \cdot \frac{z^\prime}{t^\prime}}  \, . 
\end{eqnarray}
All these equations are manifestly non-analytic.
On the world-line $z=v\cdot t$ of  ${\cal S^\prime}$ in ${\cal S}$ we have with $t\in  \mathbb{R}$:
\begin{eqnarray}  t^\prime & = &  \frac{\displaystyle t - \frac{v^\ast}{|c|^2} \cdot v\cdot t}{\displaystyle \sqrt{ 1-\left|\frac{v}{c}\right|^2}} \; = \;  t\cdot  \sqrt{ 1-\left|\frac{v}{c}\right|^2} \; \in \mathbb{R} \quad \mbox{for} \quad \left|\frac{v}{c}\right| \le 1 \; .
\end{eqnarray}
Hence, the attractive feature of a real-valued time in ${\cal S}$ and ${\cal S^\prime}$  would be obtained at the price of infering manifest non-analyticity to the theory.\\
\section{Momentum-energy covariance for complex-valued velocities}
It seems to be one of the greatest mysteries in theoretical physics that --- to our understanding --- the most straight forward derivation of Einstein's \cite{Einstein:1905xy} famous seemingly classical identity $E=mc^2$ is based on the correspondence between classical and quantum physics finding its manifestation in the concept of Louis de Broglie's \cite{deBroglie:1923xy} (1923) particle-wave duality (see also \cite{Pais2000,History2012}) in our words:\footnote{It should be stressed that the concept of ``electromagnetic mass" \cite{Pais2000,History2012} involving names like J.J.\ Thomson (1881), FitzGerald, Heaviside (1888), Searle (1896, 1897), Lorentz (1899), Wien (1900), Poincar\'{e} (1900), Kaufmann (1902-1904), Abraham (1902-1905), Hasen\"ohrl (1905)  had revealed already before Einstein (1905) a proportionality between energy and some (eventually velocity dependent) mass, i.e.\ $E\propto mc^2$. Einstein himself considered a moving body in the presence of e.m.\ radiation to derive $E=mc^2$.}  {\em In the process of quantisation the point particle of classical mechanics propagating in complex-valued space-time is replaced by energy quanta (quantum particles) being represented by some wave function $\psi$ evolving also in complex-valued space-time.} Quantum particles, i.e.\ energy quanta, can be --- depending on the spacial spread of the wave function and circumstances --- localized or delocalized. Moreover do they --- according \mbox{Liouville's} complementarity and Heisenberg's uncertainty principle --- display some simultaneous spread in complex-valued momentum space.

In the interaction-free case the wave function of a quantum particle with sharply defined momentum is a plane wave with angular frequency $\omega$ and  wave number $k$ (or wave vector $\vec{k}$ in more than one dimensions). For a real-valued space coordinate $x$ the functional behaviour of a plane wave is known to be $\psi(x,t)\propto \exp(i\,(k x-\omega t))$ yielding obviously:
\begin{eqnarray} & & \omega \; = \; + i\; \frac{\partial \ln \psi}{\partial t} \, , \qquad  k \; = \; - i\; \frac{\partial \ln \psi}{\partial x} \, .
\end{eqnarray}
As we extend our formalism to the complex plane we replace the real-valued coordinate $x$ by the complex-valued coordinate $z$ (or the complex-conjugate $z^\ast$). Instead of performing partial derivatives with respect to $x$ we will now perform partial derivatives with respect to $z$ (or $z^\ast$) which are known as  Wirtinger derivatives in one complex dimension and Dolbeault operators in several complex dimensions. They are used in the context of (anti)holomorphic functions and have the following fundamental properties being some special case of the famous Cauchy-Riemann differential equations:
\begin{eqnarray} & & \frac{\partial z^\ast}{\partial z} \; = \;  \frac{\partial z}{\partial z^\ast} \; = \; 0\, , \quad \frac{\partial z}{\partial z} \; = \;  \frac{\partial z^\ast}{\partial z^\ast} \; = \; 1\, .
\end{eqnarray}
On this formalistic ground we denote now generalized relations within a holomorphic framework to determine  the eventually complex-valued angular frequency $\omega$ and wave number $k$ for a plane wave propagating in some complexified phase space:
\begin{eqnarray} & & \omega \; = \; + i\; \frac{\partial \ln \psi}{\partial t} \, , \qquad  k \; = \; - i\; \frac{\partial \ln \psi}{\partial z} \, . \label{corrx}
\end{eqnarray}
Integration of these equations results in the following wave function for a plane wave in some some holomorphic phase space:
\begin{eqnarray} \psi(z,t) & \propto & \exp(i\,(k z-\omega t)) \, . \label{planewave}
\end{eqnarray}
As a key postulate (let's call it e.g.\ plane-wave-phase-covariance postulate (PWPCP)) in our derivation we claim at this point that {\em the eventually complex-valued phase of a plane wave should be a Lorentz scalar}. Or, in other words: {\em the eventually complex-valued phase of a plane wave should not change when boosted from one inertial frame to another}.\footnote{One could use these considerations even to define inertial frames on the basis of quantum particles: {\em An inertial frame in the absence of gravitation is some reference frame in complex-valued space-time in which the wave function describing a non-interacting quantum particle with sharply defined momentum has the mathematical form of a plane wave.}} For the previously considered inertial frames ${\cal S}$ and ${\cal S^\prime}$ this implies in particular:
\begin{eqnarray}  k z-\omega t  & = & k^\prime z^\prime -\omega^\prime t^\prime\, . \end{eqnarray}
We may now insert into the left-hand side of this equation our generalized LLFV transformations Eqs.\ (\ref{step6c}) and (\ref{step6d}), i.e.: 
\begin{eqnarray}  k \cdot \sqrt{\frac{\overline{\gamma}}{\gamma}}\, \cdot \,  \frac{\displaystyle z^\prime + v \cdot t^\prime}{\displaystyle \sqrt{ 1-\frac{v^2}{c^2}}} -\omega \cdot  \sqrt{\frac{\overline{\gamma}}{\gamma}}\, \cdot \,  \frac{\displaystyle t^\prime + \frac{v}{c^2} \cdot z^\prime}{\displaystyle \sqrt{ 1-\frac{v^2}{c^2}}} & = & k^\prime z^\prime -\omega^\prime t^\prime\, . \end{eqnarray}
Comparison of the left- and right-hand side of this equation yields the following two equations:
\begin{eqnarray} & &  k^\prime \; = \;  \sqrt{\frac{\overline{\gamma}}{\gamma}}\, \cdot \,  \frac{\displaystyle k - \frac{v}{c^2} \cdot \omega}{\displaystyle \sqrt{ 1-\frac{v^2}{c^2}}}   \, , \qquad \omega^\prime \; = \;  \sqrt{\frac{\overline{\gamma}}{\gamma}}\, \cdot \,  \frac{\displaystyle \omega - v \cdot k}{\displaystyle \sqrt{ 1-\frac{v^2}{c^2}}} \, , \label{stepxa} 
\end{eqnarray}
and by application of the principle of relativity interchanging $k$, $\omega$  and $k^\prime$, $\omega^\prime$, respectively, and replacing $v$ by $-v$ the two inverse equations:
\begin{eqnarray}
 & &   k \; = \;  \sqrt{\frac{\gamma}{\overline{\gamma}}}\, \cdot \,  \frac{\displaystyle k^\prime + \frac{v}{c^2} \cdot \omega^\prime}{\displaystyle \sqrt{ 1-\frac{v^2}{c^2}}}  \, , \qquad \omega \; = \;  \sqrt{\frac{\gamma}{\overline{\gamma}}}\, \cdot \,  \frac{\displaystyle \omega^\prime + v \cdot k^\prime}{\displaystyle \sqrt{ 1-\frac{v^2}{c^2}}}  \, . \label{stepxb}
\end{eqnarray}
These four equations state that an eventually complex-valued frequency $\omega$ and some ---  3-dimensionally generalized ---  wave vector $\vec{k}$ are transforming like a four vector under {\em inverse} generalized LLFV transformations.

While the wave representation of particles has brought us to the quantum formalism without even involving Planck's quantum of action  $\hbar=h/(2\pi)=1.054571726(47)\cdot 10^{-34}$~J$\,$s \cite{mohr2012} it is the following two highly non-trivial and fundamental identities conjectured by Louis de Broglie \cite{deBroglie:1923xy} (1923) to be applicable even to {\em massive} particles which will bring us back to the seemingly classical quantities momentum $p$ (here for simplicity in one  dimension) and energy $E$, i.e. (with $k=2\pi/\lambda$):\footnote{It is of course known that the former identity $E= h f$ (with $f=\omega/(2\pi)$) had been derived earlier --- using an energy-discretisation trick of  Boltzmann  --- by Planck (1900)  in the context of the e.m.\ radiation of a black body  and by Einstein (1905) to determine the energy of his {\em massless} photon, while the latter identity had been used in the form $p=hf/c$ for {\em massless} photons for the first time by Stark (1909) and later by Einstein (1916, 1918) while Compton (1923) and Debye (1923)  had finally confirmed the proportionality of the suggested three-momentum and wave vector of a {\em massless} photon by famous experiments.}
\begin{eqnarray} & & p \; = \; \hbar \, k \, , \quad E \; = \; \hbar\,\omega \, \label{stepdb}
\end{eqnarray}
and --- when combined with our Eqs.\ (\ref{corrx}) --- to the following two fundamental identies representing even for wave functions of interacting quantum particles (being not of plane wave form) the correspondence principle of QHJT, i.e.:
\begin{eqnarray} & & E \; = \; + i\, \hbar \; \frac{\partial \ln \psi}{\partial t} \, , \qquad  p \; = \; - i\, \hbar\; \frac{\partial \ln \psi}{\partial z} \, . \label{corry}
\end{eqnarray}
In multiplying Eqs. (\ref{stepxa}) and (\ref{stepxb}) by $\hbar$ it is now straight forward to obtain via Eqs.\ (\ref{stepdb}) the seemingly classical generalized Lorentz-Planck (LP) transformations (see also Planck \cite{Planck:1906xy} (1906)) relating here some even eventually complex-valued momentum and energy in inertial frames ${\cal S}$ and ${\cal S^\prime}$:\footnote{Without loss of generality we display the equations here only for one complex-valued momentum dimension.} 
\begin{eqnarray} & &  p^\prime \; = \;  \sqrt{\frac{\overline{\gamma}}{\gamma}}\, \cdot \,  \frac{\displaystyle p - \frac{v}{c^2} \cdot E}{\displaystyle \sqrt{ 1-\frac{v^2}{c^2}}}   \, , \qquad E^\prime \; = \;  \sqrt{\frac{\overline{\gamma}}{\gamma}}\, \cdot \,  \frac{\displaystyle E - v \cdot p}{\displaystyle \sqrt{1-\frac{v^2}{c^2}}} \, , \label{stepya} \\[2mm]
 & &   p \; = \;  \sqrt{\frac{\gamma}{\overline{\gamma}}}\, \cdot \,  \frac{\displaystyle p^\prime + \frac{v}{c^2} \cdot E^\prime}{\displaystyle \sqrt{ 1-\frac{v^2}{c^2}}}  \, , \qquad E \; = \;  \sqrt{\frac{\gamma}{\overline{\gamma}}}\, \cdot \,  \frac{\displaystyle E^\prime + v \cdot p^\prime}{\displaystyle \sqrt{ 1-\frac{v^2}{c^2}}}  \, . \label{stepyb}
\end{eqnarray}
Hence, a particle with zero momentum  ($p^\prime=0$) in ${\cal S^\prime}$ will have in the frame ${\cal S}$ of a resting observer the eventually complex-valued velocity $v$ and appear with eventually complex-valued momentum $p$ and energy $E$ given  by:\footnote{In the limit $\gamma =  \overline{\gamma}$ we recover the famous relativistic identities $\vec{p} = m \vec{v}$ (Planck \cite{Planck:1906xy} (1906)) and $E=mc^2$ (Einstein \cite{Einstein:1905xy} (1905)) with $m=\frac{m_0}{\sqrt{1-(v/c)^2}}$.} 
\begin{eqnarray} & &   p \; = \;   \sqrt{\frac{\gamma}{\overline{\gamma}}}\, \cdot \,  \frac{\displaystyle m_0 \, v}{\displaystyle \sqrt{1-\frac{v^2}{c^2}}} \, , \quad
 E \; = \;  \sqrt{\frac{\gamma}{\overline{\gamma}}}\, \cdot \,  \frac{\displaystyle m_0 \, c^2}{\displaystyle \sqrt{ 1-\frac{v^2}{c^2}}} \, , \label{stepzb}
\end{eqnarray}
with $m_0\equiv E^\prime/c^2$ being some eventually complex-valued rest mass and Lorentz invariant as there obviously holds the generalized dispersion relation:
\begin{eqnarray} E^2 - (pc)^2 & = & \frac{\gamma}{\overline{\gamma}} \; (m_0\,c^2)^2\, .\label{disprel1} \end{eqnarray}
\section{Non-Hermitian Klein-Gordon-Fock equation}
At this point we would like to recall Eqs.\ (\ref{corry}) expressing  the correspondence principle in QHJT and being even valid for {\em interacting} quantum particles:
\begin{eqnarray} & & E \; = \;  + i\, \hbar\; \frac{\partial \psi}{\partial t} \,\cdot \, \frac{1}{\psi} \, , \qquad  p \; = \; - i\, \hbar\; \frac{\partial  \psi}{\partial z} \,\cdot \, \frac{1}{\psi} \, , \label{corrz}
\end{eqnarray}
yielding obviously
\begin{eqnarray} & & E^2 \; = \;  (+ i\, \hbar)^2\; \left(\frac{\partial \psi}{\partial t}\right)^2 \,\cdot \, \frac{1}{\psi^2} \, , \qquad  p^2 \; = \; (- i\, \hbar)^2\; \left(\frac{\partial  \psi}{\partial z}\right)^2 \,\cdot \, \frac{1}{\psi^2} \, , \label{corrzz}
\end{eqnarray}
Simultaneously there are the following two identities holding for {\em non-interacting} quantum particles being described by a plane wave $\psi \propto \exp\left(\frac{i}{h} (p\, z - E\, t)\right)$ (obtained by combining Eq.\ (\ref{planewave}) with Eqs.\ (\ref{stepdb})):
\begin{eqnarray} & & E^2 \; = \;  (+ i\, \hbar)^2\; \frac{\partial^2 \psi}{\partial t^2} \,\cdot \, \frac{1}{\psi} \, , \qquad  p^2 \; = \; (- i\, \hbar)^2\; \frac{\partial^2  \psi}{\partial z^2} \,\cdot \, \frac{1}{\psi} \, , \label{corrzz}
\end{eqnarray}
As Klein \cite{Klein:1926tv}, Gordon \cite{Gordon:1926xy} and Fock \cite{Fock:1926fj} in 1926 we can insert Eqs.\ (\ref{corrzz}) in the dispersion relation Eq.\  (\ref{disprel1}) to obtain the generalized Klein-Gordon-Fock (KGF) equation describing a {\em non-interacting} relativistic quantum particle:
\begin{eqnarray} & & \overbrace{(+ i\, \hbar)^2\; \frac{\partial^2 \psi}{\partial t^2} \,\cdot \, \frac{1}{\psi}}^{\displaystyle E^2} -  \overbrace{(- i\, \hbar\,c)^2\; \frac{\partial^2  \psi}{\partial z^2} \,\cdot \, \frac{1}{\psi}}^{\displaystyle (p c)^2} \; = \; \frac{\gamma}{\overline{\gamma}} \; (m_0\,c^2)^2 \label{KG1} \\[3mm]
 & \Longrightarrow & (+ i\, \hbar)^2\; \frac{\partial^2 \psi}{\partial t^2} \, - \, (- i\, \hbar\,c)^2\; \frac{\partial^2  \psi}{\partial z^2}  \; = \; \frac{\gamma}{\overline{\gamma}} \; (m_0\,c^2)^2\; \psi \label{KG2} \\[3mm] 
 & \Longrightarrow & (+ i\, \hbar)^2\; \frac{\partial^2 \psi}{\partial t^2} \; =\; (- i\, \hbar\,c)^2\; \frac{\partial^2  \psi}{\partial z^2}  +  \frac{\gamma}{\overline{\gamma}} \; (m_0\,c^2)^2\; \psi \, . \label{KG3} 
\end{eqnarray}
As usual the solution $\psi = \psi^{(+)} + \psi^{(-)}$ of the KGF Eq.\ (\ref{KG3}) can be decomposed into a sum of a retarded solution $\psi^{(+)}$ and an advanced solution $\psi^{(-)}$ solving not only the KGF Eq.\ (\ref{KG3}), yet also respectively the following relativstic retarded or advanced interaction free Schr\"odinger \cite{Schrodinger:1926xy} (1926) equations:
\begin{eqnarray}   \pm \,  i\, \hbar\; \frac{\partial \psi^{(\pm)}}{\partial t} & = & \sqrt{(- i\, \hbar\,c)^2\; \frac{\partial^2}{\partial z^2}  +  \frac{\gamma}{\overline{\gamma}} \; (m_0\,c^2)^2}\;\; \psi^{(\pm)} \label{SE1} \\[2mm]
 & \approx & \left( \sqrt{ \frac{\gamma}{\overline{\gamma}}}  \; m_0\,c^2 -\, \sqrt{ \frac{\overline{\gamma}}{\gamma}} \; \frac{\hbar^2}{2\,m_0}\; \frac{\partial^2}{\partial z^2} + \ldots \right) \; \psi^{(\pm)}  \label{SE2}
\end{eqnarray}
In the last line we performed the non-relativistic limit well known for $\gamma =\overline{\gamma}$. 
\section{Non-Hermitian Dirac-equation}
Each of the four components of the Dirac spinor $\psi$ of a {\em non-interacting} Dirac-quantum particle should individually respect the  KGF Eq.\ (\ref{KG1}). Returning to three eventually complex-valued space and momentum dimensions this condition is formally denoted by the following equivalent identities:
\begin{eqnarray} 0 & = & \left( E^2 - (\vec{p}\, c)^2 -  \frac{\gamma}{\overline{\gamma}} \; (m_0\,c^2)^2 \right) \psi \label{dirac1} \\[1mm]
  \Rightarrow \quad 0 & = & \left((+ i\, \hbar)^2\; \frac{\partial^2}{\partial t^2}  - (- i\, \hbar\,c)^2\; \frac{\partial}{\partial \vec{z}} \cdot \frac{\partial}{\partial \vec{z}}  -  \frac{\gamma}{\overline{\gamma}} \; (m_0\,c^2)^2 \right) \psi \label{dirac2} \\[1mm]
  \Rightarrow \quad 0 & = & \left(\left[\, \beta \left( + i\, \hbar\,  \frac{\partial}{\partial t}  - (- i\, \hbar\,c)\; \vec{\alpha} \cdot  \frac{\partial}{\partial \vec{z}} \right) \right]^2  -  \frac{\gamma}{\overline{\gamma}} \; (m_0\,c^2)^2 \right) \psi \label{dirac3} \\[1mm]
  \Rightarrow \quad 0 & = &  \left(\, \beta \left( + i\, \hbar\,  \frac{\partial}{\partial t}  - (- i\, \hbar\,c)\; \vec{\alpha} \cdot  \frac{\partial}{\partial \vec{z}} \right)   -  \sqrt{\frac{\gamma}{\overline{\gamma}}} \; m_0\,c^2 \right)  \nonumber   \\[1mm]
 &  & \left(\, \beta \left( + i\, \hbar\,  \frac{\partial}{\partial t}  - (- i\, \hbar\,c)\; \vec{\alpha} \cdot  \frac{\partial}{\partial \vec{z}} \right)   +  \sqrt{\frac{\gamma}{\overline{\gamma}}} \; m_0\,c^2 \right) \psi  \, . \label{dirac4} \end{eqnarray}
Throughout the factorization of Eq.\ (\ref{dirac2}) we made use of the four well known $4\times 4$ Dirac matrices $\vec{\alpha}$ and $\beta$, which are defined as follows with the help of the Pauli matrices $\vec{\sigma}$, the $2\times 2$ unit matrix $1_2$ and the $2\times 2$ zero matrix $0_2$:
\begin{eqnarray} & & \vec{\alpha} \; \equiv \; \left( \begin{array}{cc} 0_2 & \vec{\sigma} \\ \vec{\sigma} & 0_2 \end{array}\right) \; , \qquad \beta \equiv  \left( \begin{array}{cc} 1_2 & 0_2 \\ 0_2 & - 1_2 \end{array}\right) \; . 
\end{eqnarray}
By simple inspection of Eq.\ (\ref{dirac4}) and use of the identiy $\beta^2=1_4$ it is now straight forward to denote the retarded and advanced Dirac \cite{Dirac:1928hu} (1928) equations for the retarded component $\psi^{(+)}$ and advanced component $\psi^{(-)}$ of solution $\psi=\psi^{(+)}+\psi^{(-)}$ of the {\em interaction free} KGF Eq.\ (\ref{dirac1}), i.e.:  
\begin{eqnarray}
 0 & = & \left(\, \beta \left( + i\, \hbar\,  \frac{\partial}{\partial t}  - (- i\, \hbar\,c)\; \vec{\alpha} \cdot  \frac{\partial}{\partial \vec{z}} \right)   \mp  \sqrt{\frac{\gamma}{\overline{\gamma}}} \; m_0\,c^2 \right) \psi^{(\pm)}  \, \\[1mm]
 & \Rightarrow & + i\, \hbar\,  \frac{\partial \psi^{(\pm)}}{\partial t} \; = \; \left( - i\, \hbar\,c\; \vec{\alpha} \cdot  \frac{\partial}{\partial \vec{z}} \; \pm \sqrt{\frac{\gamma}{\overline{\gamma}}} \; \beta \;  m_0\,c^2 \right) \psi^{(\pm)}  \, \\[1mm]
 & \Rightarrow & \pm\,  i\, \hbar\,  \frac{\partial \psi^{(\pm)}}{\partial t} \; = \; \left( \pm \, (- i\, \hbar\,c)\; \vec{\alpha} \cdot  \frac{\partial}{\partial \vec{z}} \; + \sqrt{\frac{\gamma}{\overline{\gamma}}} \; \beta \;  m_0\,c^2 \right) \psi^{(\pm)}  \, . \label{dirac5} 
\end{eqnarray}
Once more we stress that these generalized Dirac equations, the generalized Schr\"odinger Eqs.\ (\ref{SE1}), (\ref{SE2}) and the generalized KGF Eqs.\ (\ref{KG3}), (\ref{dirac2}) do hold even in complex-valued space-time and for complex-valued rest mass $m_0$.
\section{Final remarks}
The purpose of the presented considerations has been to extend the concept of covariance to complex-valued space time. It is remarkable that this can be achieved in some analytical way on the basis and in accordance with the correspondence principle of QHJT. After extending the concept of inertial frames to the complex plane we have constructed on one hand generalized  LLFV and LP transformations relating the four vectors of complex-valued space-time and momentum-energy beween two inertial frames with an eventually  complex-valued relative velocity, on the other hand  a complex generalization of Einstein's energy-mass equivalence $E = m\,c^2$. It turned out that the complexification of time is not a severe problem as a boost will multiply the time at most by a complex constant. Moreover it has been possible to derive on the basis of a generalized concept of covariance generalized KGF, Schr\"odinger and Dirac differential equations which can be used to formulate of a non-Hermitian quantum theory describing the apparently complex laws of physics. As had been pointed out already earlier (e.g.\ \cite{Kleefeld:2004jb}) it is the advanced  Schr\"odinger (or Dirac) equation which plays the role of Benders hardly constructable CPT-transformed Schr\"odinger (or Dirac) equation required to construct some positive semidefinite CPT-inner product \cite{Bender:2002vv} for some PT-symmetric quantum theory. The possibility to obtain via covariance directly the underlying advanced Schr\"odinger (or Dirac) equation as described in the present paper will make the tedious search and construction of a unique CPT-inner product in non-Hermitian quantum theory needless. 

At this place we would like to thank Miloslav Znojil for the strong encouragement to publish the present --- already comprehensive --- results despite the fact that --- to our understanding --- there remain still many questions open to be answered elsewhere in future.


\subsection*{Acknowledgements}

This work was supported by the Funda\c{c}\~{a}o para a Ci\^{e}ncia e a Tecnologia of the 
Minist\'{e}rio da Ci\^{e}ncia, Tecnologia e Ensino Superior of Portugal, under contract
CERN/FP/123576/2011  and by the Czech project
LC06002.


\end{document}